# Predicting drug recalls from Internet search engine queries

Elad Yom-Tov

Microsoft Research, Israel

## Abstract

Batches of pharmaceutical are sometimes recalled from the market when a safety issue or a defect is detected in specific production runs of a drug. Such problems are usually detected when patients or healthcare providers report abnormalities to medical authorities. Here we test the hypothesis that defective production lots can be detected earlier by monitoring queries to Internet search engines.

We extracted queries from the USA to the Bing search engine which mentioned one of 5,195 pharmaceutical drugs during 2015 and all recall notifications issued by the Food and Drug Administration (FDA) during that year. By using attributes that quantify the change in query volume at the state level, we attempted to predict if a recall of a specific drug will be ordered by FDA in a time horizon ranging from one to 40 days in future.

Our results show that future drug recalls can indeed be identified with an AUC of 0.791 and a lift at 5% of approximately 6 when predicting a recall will occur one day ahead. This performance degrades as prediction is made for longer periods ahead. The most indicative attributes for prediction are sudden spikes in query volume about a specific medicine in each state. Recalls of prescription drugs and those estimated to be of medium-risk are more likely to be identified using search query data.

These findings suggest that aggregated Internet search engine data can be used to facilitate in early warning of faulty batches of medicines.

## 1  Introduction

A drug recall occurs when a batch or an entire production run of a drug product is returned to the manufacturer, usually due to the detection of safety issues or drug product defect [1]. Drug recalls are costly for manufacturers, in both direct costs – loss of sales and cost of collecting the faulty drug -- and indirect ones, such as loss of goodwill [2].

Here we focus on recalls of specific batches of drugs, not of entire drug recalls. This is because the former are relatively common (as we show below, 3772 recalls were logged in 2015), compared to entire drug recalls, which have not occurred in the US since 2011. Whether initiated by the manufacturer or by the Food and Drug Agency (FDA) in the United States, recalls are logged by the FDA and provided to the public via the FDA's Drug Recall Enforcement Reports Reference[1].

Internet data, including social media posts [3] and search engine queries, have been used previously to identify adverse reactions of medical drugs [4], [5]. For example, Yom-Tov and Gabrilovich [6] showed that queries to Internet search engines can be used to monitor and detect possible adverse reactions of medicines. Broadly, people are likely to query for drugs when these are prescribed to them, and for adverse reactions when they are experienced by them. By comparing the number of people who queried for a drug and later queried for specific adverse reactions, compared to other people, it is possible to identify candidate adverse reactions. These have been shown to match known adverse reaction of pharmaceutical drugs, as well as unknown reactions, which share the trait of being more benign and appear after a longer time than of known adverse reactions.

Recent analysis has found that the most common reasons for drug recalls are contamination, mislabeling, adverse reaction, defective product, and incorrect potency [7]. Since some of the reasons for recalls may be experienced by the consumer as causing adverse reactions or ineffective products, we hypothesize that mechanisms similar to those used for detection of general adverse reactions will be effective in early detection of faulty medicines, which will later be cause for recall. Specifically, we propose to use changes in the query volume for drugs and adverse reactions as an indicator for the possible existence of faulty drug batches.

Thus, we perform a large-scale retrospective analysis of Internet search engine data and show that these data can indeed be used as a sentinel for detecting faulty batches of pharmaceutical drugs.

---

[1] https://open.fda.gov/drug/enforcement/reference/

## 2 Methods

### 2.1 Data

We extracted all queries submitted to the Bing search engine by users in the United States between January 1st, 2015 and December 31st, 2015. For each query were recorded an anonymized user identifier, the text of the query, the date when the query was made, and the US state where the user was situated when issuing the query.

All procedures performed in this study were in accordance with the ethical standards of the institutional. For this type of study formal consent is not required.

Queries were filtered to include those whose text contained one or more of 5,195 drugs listed in Wikipedia, either in their generic or brand names. We marked queries as to whether they contained a medical symptom, according to the list compiled in Yom-Tov and Gabrilovich [6]. Drugs were filtered to keep 373 drugs for which at least 1000 queries were made in the data.

The ground truth which we attempted to identify were the recalls listed in the US Food and Drug Administration (FDA) Recall Enterprise System (RES), a database that contains information on recall event information submitted to FDA. Recalls in RES are listed by the states affected by the recall, the date of the recall, and the drug that was recalled. Also available in these data are the recall classification, where a Class I recall is due to "Dangerous or defective products that predictably could cause serious health problems or death", Class II to "Products that might cause a temporary health problem, or pose only a slight threat of a serious nature" and Class III to "Products that are unlikely to cause any adverse health reaction, but that violate FDA labeling or manufacturing laws".

Finally, we classified drugs as to whether they require a prescription (RX) or are sold without one (OTC) using the FDA's Orange Book [8].

### 2.2 Data modeling

We assumed that faulty drugs would manifest as changes in the query volume about these medications, compared to the normal volume of queries about these drugs. Thus, for each day, state and drug combination we computed the following 20 time-series attributes:

1. **Drug query slope**: Slope of the number of queries about the drug that were made in the state during the past week, 2 weeks, and up to 7 weeks.
2. **Drug-symptom query slope**: The same as (1), but only for queries that also mention a symptom.
3. **Drug spike ratio 1/7**: The relationship between the number of queries about the drug in the state during the past day and the same number in the past 7 days.
4. **Drug spike ratio 1/30**: The same as (3) for the ratio between the past day and the past 30 days.
5. **Drug spike ratio 7/30**: The same as (3) for the ratio between the past 7 days and the past 30 days.

6. **Drug-symptom spike ratio 1/7**: The same as (3), but only for queries that also mention a symptom.
7. **Drug-symptom spike ratio 1/30**: The same as (4), but only for queries that also mention a symptom.
8. **Drug-symptom spike ratio 7/30**: The same as (5), but only for queries that also mention a symptom.

All data for a given drug in a state was removed from the day of the first recall up to the end of the year, so that publicity and media coverage of a recall would not affect our estimation of the ability to detect a recall.

## 2.3 Prediction

We attempted to predict if a recall would be ordered for each of the 5,195 drugs we monitored in each of the 50 US states, N days in the future. We modified N between 1 (the call would be ordered in the next day) to 40 days (a recall would be ordered in 40 days). We refer to *N* as the *predictive horizon*. Once the date of a recall was passed for a given drug in each state, all future instances of this drug would be ignored in this state, so that any public health information about the recall or media attention given to the recall would not be taken into account in evaluation.

Data were split into train and test, such that the first 240 days of 2015 were used as training data, and the last 85 days as test data. Such stratification (by time) mimics the way a system could be deployed in production, where historical data would be used for training, and current data would be analyzed for potential recalls.

The fraction of positive samples (recalls) was approximately 0.2%. Thus, it is necessary to use prediction methods that account for rare samples in the predicted class. Here we employ a version of bagging [9], which (following [10]) works as follows: First, the majority class (no recall) are clustered using the k-means algorithm. We determined empirically that good results are obtained for k=500. Then, a linear predictor with interactions is constructed to distinguish between the examples in each cluster and the positive examples of the training set. During prediction, each of the classifiers is applied to the example, and the label is set to be the maximum value outputted by the predictors.

We evaluated the performance of our prediction using two measures: Receiver Operating Curve (ROC) and the corresponding Area Under the Curve (AUC), and lift. ROC (and AUC) represent the balance between false positive and true positive rates of a classifier and is one of the most common measures for evaluation of classifiers. Lift [11], [12], for any given fraction $0 < T < 1$, is defined as the ratio between the number of positive examples among the fraction of T examples, that are ranked highest by the proposed predictor, and the expected number of positive examples in a random sample from the general set of samples of equal size. For example, a lift of 3 at a fraction T = 0.05 means that if we examined 1% of drugs at states in a given data ranked highest by the proposed system, we expect to see three times more drugs which will require to be recalled in this population than in a 0.05-fraction random sample of the examples.

However, two reasons suggest that lift should also be used in our evaluation. First, the sparsity of our recalled class (0.2%) means that small variations in the size of this class may cause large deviations in the observed performance (See, for example, [13] and citations therein). Second, in a practical scenario, regulators are likely to prefer an ordered list of drugs that should be inspected (from top to bottom), as budget and time allows. Therefore, results for both measures of performance are given.

## 3   Results

On average, there were 3772 recalls in 2015. Figure 1 shows the number of recalls in each state, excluding 3462 recalls which were applied to all states. As can be seen, more populous states experienced more recalls, as might be expected. However, even given this large number of recalls, given that we are attempting to identify recalls before they occur, a typical system would only experience approximately 0.2% positive examples (3772 recalls per state, compared to 5192 drugs in 365 days).

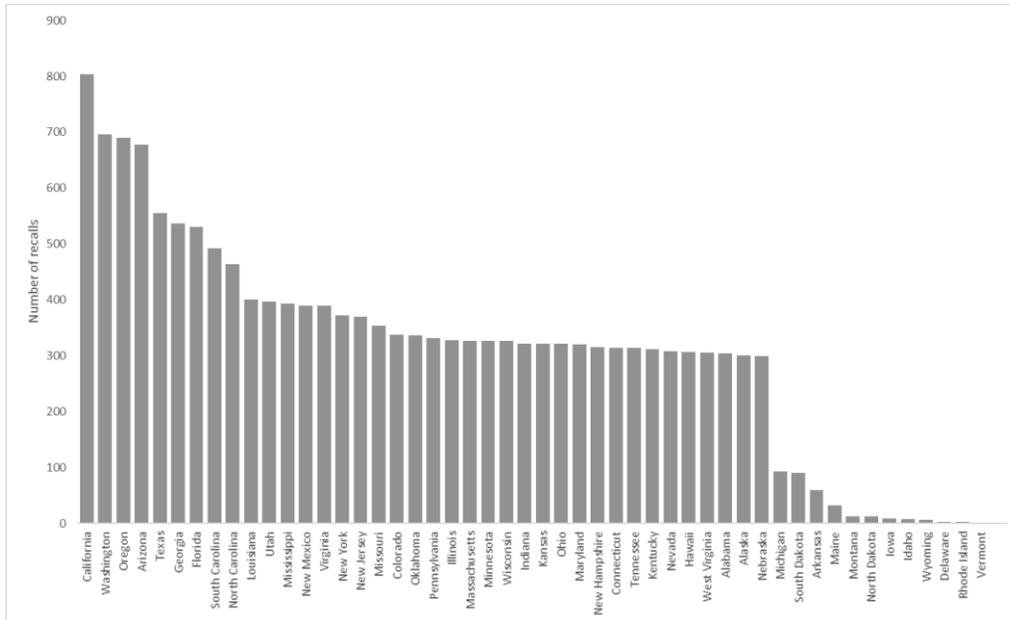

*Figure 1: Number of non-US-wide recalls per state reported by FDA during 2015*

Figure 2 shows the ROC (with AUC of 0.791) and the lift chart for predicting recalls one day before they occur. As this figure show, it is possible to identify around 20% of the recalls at a relatively low false positive rate.

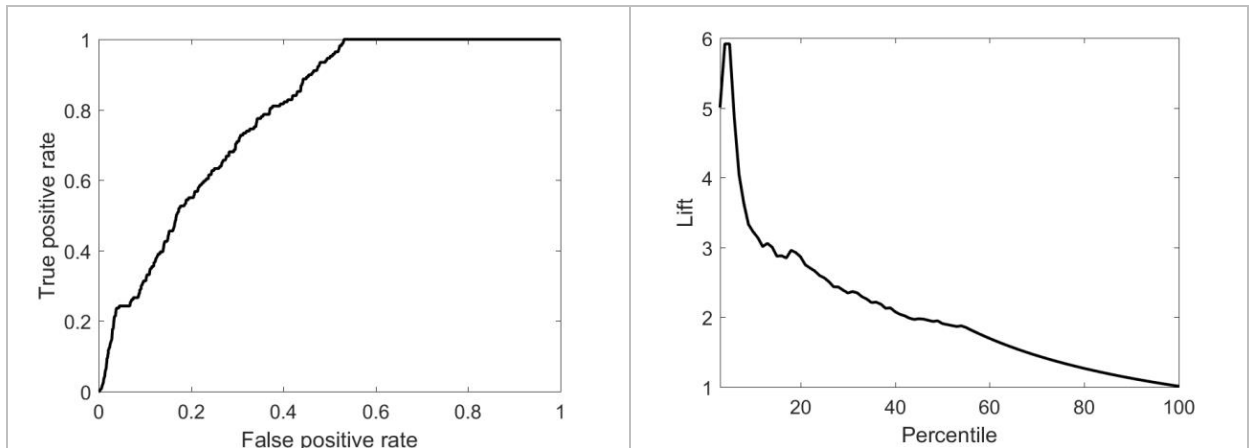

*Figure 2: ROC (left) and lift (right) for detecting recalls one day before they occur.*

Figure 3 shows the relationship between the number of days before recall that a prediction is made and lift at 5%. As expected, the longer into the future that the predictor attempts to identify a likely recall, the worst the performance. However, there is a large variance in performance among adjacent points. To explain this variance, we modeled the performance of the classifier (both using AUC and lift at 5%) as a function of two parameters: The number of days before recall, and the number of positive examples in the test data. The model applied to these variables is a rank regression model.

The resulting model parameters are shown in Table 1, from which several observations can be made. First, the model cannot explain the variation in AUC, whereas variations in lift are well explained, with $R^2$ for the former being 0.13 (not statistically significant) and 0.46 (P=0.02) for the latter. We attribute this difference to the sensitivity of AUC to small variations in performance, as noted above. Second, for the model of lift, performance degrades as a function of time before recall (negative slope), and as fewer positive examples exist (positive slope). Thus, the variance in Figure 2 can be explained largely as a result of two effects: The difficulty of predicting long into the future (a parameter of the task), and the sparsity of the data (a parameter of the test data).

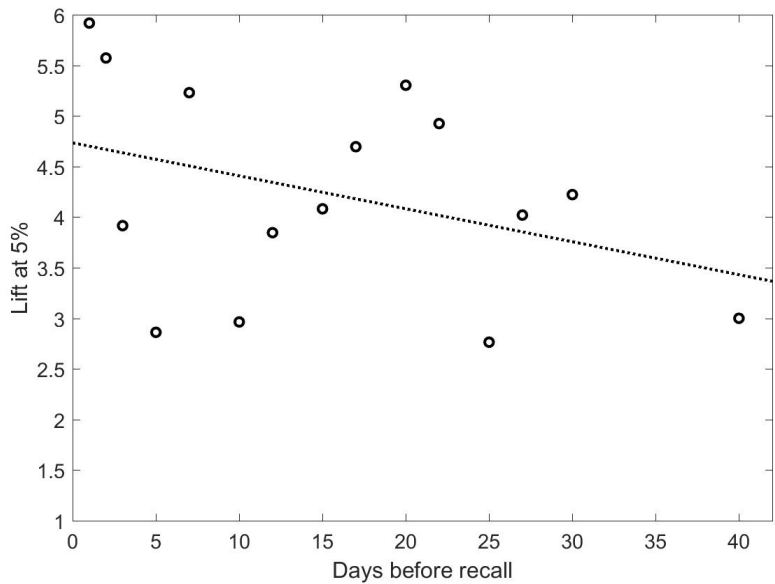

*Figure 3: Lift at 5% as a function of the number of days before recall that prediction is made. The dotted line represents the best fit linear regression line.*

|  | AUC | | Lift @ 5% | |
|---|---|---|---|---|
|  | Slope (SE) | p-value | Slope (SE) | p-value |
| **Time to recall** | -1.25 (0.92) | 0.201 | -2.31 (0.73) | 0.008 |
| **Number of positive examples** | 1.25 (0.93) | 0.204 | 2.11 (0.73) | 0.013 |
| **$R^2$** | 0.13 | 0.422 | 0.46 | 0.024 |

*Table 1: Performance as a function of time to recall and the number of positive examples*

Recalls ranked in the top 5% per the prediction model for a horizon of one day were stratified by their recall class (see Methods) and whether the drug recalled required a prescription (RX) or was sold over the counter (OTC). We compared the likelihood of each recall class and (separately) whether the drug was RX or OTC to the likelihood in the entire set of recalls. Recalls of class II (medium danger) were 24% more likely to appear in the top 5% of the predictions. RX medications were 14% more likely while OTCs and unclassified drugs were 69% less likely to appear in the top 5% of the predictions. Thus, it was easier to detect recalls that are of medium danger and recalls of prescription drugs.

As explained in the Methods, the classifier is an ensemble classifier based on 500 clusters of the (majority) negative class. There is a strong correlation (Spearman rho 0.6, $P<10^{-6}$) between the number of points in each cluster and the number of times that the output of the classifier based on points from a cluster are used for classification (because their value is the largest among classifiers). Based on this insight, we measured the maximal lift obtained by the ensemble, when only the largest clusters are used. Figure 4 shows the maximal lift for predicting recalls one day before they occur, as a function of the number of clusters used, starting from the single largest cluster and up to the 100 largest clusters.

Similar graphs are observed for other predictive horizons. As the figure shows, performance increases when between approximately 5 and 60 of the largest clusters are used. However, this increase in performance is non-monotonous. Therefore, it may be beneficial to use only a number of the largest clusters when using the ensemble, but the best number of chosen clusters to be used is difficult to determine without empirical testing.

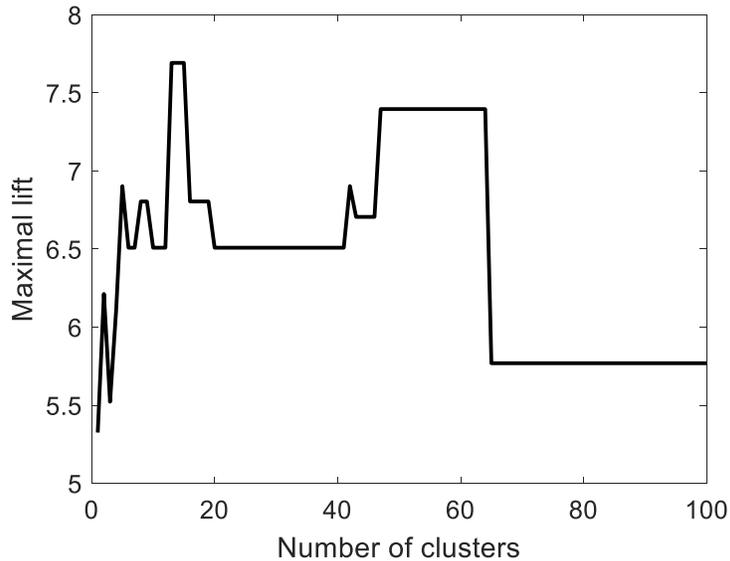

*Figure 4: Classifier performance as a function of the number of clusters used, when using consecutively smaller clusters*

It is also possible to estimate the contribution of different attributes to the ensemble classifier. This is done by finding those variables that appear as statistically significant (at $P<0.05$, with Bonferroni correction) in a significant fraction of the 500 classifiers. If we choose to focus on attributes that appear (either independently or in an interaction with other attributes) as significant in at least 20% of the classifiers, we find that the most influential attributes are (see Methods for a description of the attributes):

1. Drug query slope.
2. Drug spike ratio 1/7
3. Drug spike ratio 1/30
4. Drug spike ratio 7/30

Similar influential attributes are observed for other predictive horizon values. Thus, changes, and especially sudden ones, in queries for a drug are indicative of a pending recall.

## 4  Discussion

Early detection of faulty batches of pharmaceutical drugs is important for both patients and pharmaceutical companies. For the former, such detection reduces the risk of adverse events and ineffective treatment. For the latter, it can assist in rapid reaction to manufacturing problems which can result in significant financial consequences if untreated.

Internet data, which reflect patient's experience in using drugs, has been shown to be beneficial for the discovery of adverse reactions. This is achieved through monitoring of very large populations of Internet users as they query about their experiences or discuss them on Internet forums. Here we have shown the possibility to monitor, in near real-time, Internet data to aid in early discovery of specific batches of faulty pharmaceuticals.

One of the main problems encountered when predicting rare events such as drug recalls, and when evaluating the success of such prediction, is the scarcity of such recalls. Our results show that prediction quality depends on how close to the recall we try to detect it, and the number of recalls in our testing data. Because of the latter, our analysis should be regarded as showing the feasibility for detecting drug recalls. Future research will need to focus on analyzing data for longer timeframes, thus providing a more accurate predictor and a more stable estimation of classification accuracy.

We found that changes, and especially sudden ones, in queries for a drug are indicative of a pending recall. Interestingly, the confluence of symptoms and drugs were not found to be strongly indicative of drug recalls. It may be that better attributes could be developed by counting the number of people who searched for each drug (ostensibly because they or someone close to them were prescribed this drug) and later searched for an adverse reaction. However, such attributes are more difficult to compute and are less available to public health authorities, compared to the attributes we used, which can be collected, for example, through services such as Google Trends[2]. Thus, the use of these attributes is left for future research.

---

[2] www.google.com/trends

# 5 References

1. Nagaich, U., Sadhna, D.: Drug recall: An incubus for pharmaceutical companies and most serious drug recall of history. International Journal of Pharmaceutical Investigations 5(1), 13-19 (2015)

2. Jarrell, G., Peltzman, S.: The Impact of Product Recalls on the Wealth of Sellers. Journal of Political Economy 93(3), 512-536 (1985)

3. Yates, A., Goharian, N.: ADRTrace: detecting expected and unexpected adverse drug reactions from user reviews on social media sites. In : European Conference on Information Retrieval, pp.816-819 (2013)

4. Yom-Tov, E.: Crowdsourced Health: How What You Do on the Internet Will Improve Medicine. MIT PRess, Boston, MA (2016)

5. Pages, A., Bondon-Guitton, E., Montastruc, J., Bagheri, H.: Undesirable effects related to oral antineoplastic drugs: comparison between patients' internet narratives and a national pharmacovigilance database. Drug Safety 37(8), 629-637 (2014)

6. Yom-Tov, E., Gabrilovich, E.: Postmarket drug surveillance without trial costs: discovery of adverse drug reactions through large-scale analysis of web search queries. Journal of medical internet research 15(6), e124 (2013)

7. Hall, K., Stewart, T., Chang, J., Freeman, M.: Characteristics of FDA drug recalls: A 30-month analysis. American Journal of Health-System Pharmacy 73(4), 235-240 (2016)

8. Food and Drug Administration: Approved Drug Products, 36th Edition. (2016)

9. Mikel Galar, A.: A Review on Ensembles for the Class Imbalance Problem: Bagging-, Boosting-, and Hybrid-Based Approaches. IEEE TRANSACTIONS ON SYSTEMS, MAN, AND CYBERNETICS—PART C: APPLICATIONS AND REVIEWS 42, 463-484 (July 2012)

10. Jo, T., Japkowicz, N.: Class imbalances versus small disjuncts. ACM SIGKDD Explorations Newsletter 6(1), 40-49 (2004)

11. Richter, Y., Yom-Tov, E., Slonim, N.: Predicting Customer Churn in Mobile Networks through Analysis of Social Groups. In : SDM, pp.732-741 (2009)

12. Lima, E.: Domain knowledge integration in data mining for churn and customer lifetime value modelling: new approaches and applications. (2009)

13. Hanczar, B., Hua, J., Sima, C., Weinstein, J., Bittner, M., Dougherty, E.: Small-sample precision of ROC-related estimates. Bioinformatics 26(6), 822-830 (2010)